\documentclass{article}
\usepackage{graphicx}
\begin{document}
\title{Pressure-temperature phase diagrams of selenium and sulfur
 in terms of Patashinski model}%
\author{L.Son$^{1}$, G.Rusakov$^{2}$, N.Katkov$^{1}$}
\maketitle
%\institute
$^{1}$Ural State Pedagogical University, 620219 Ekaterinburg,
Russia.e-mail:ldson@mail.ru\\
 $ ^{2}$Institute for metal physics, Ural division of Russian
  academy of sciences, 620000 Ekaterinburg, Russia.

%\end{opening}

\begin{abstract}
The pressure - temperature phase diagrams of {\bf Se} and {\bf S}
are calculated. Both melting and polymorphous phase transitions
are described  in  frames of statistical Patashinski model. The
results are in good agreement with experimental data of Brazhkin
et. al.
\end{abstract}

{\bf keywords:}~Patashinski model, phase diagram, selenium,
sulfur.

\section{Inroduction.}

Structural transitions in liquid {\bf Se} and {\bf S} similar to
those in crystalline state were discovered by Brazhkin {et.al.}
\cite{b1}. At these transitions, drastic changes of structure and
physical properties (electrical conductivity, heat capacity, etc.)
occur
\cite{b2}. Actually, the pressure - temperature phase diagrams of
these substances demonstrate two kinds of phase transitions. The
first one corresponds to the sharp change of local order in the
system. In solid state, these transitions are known as
polymorphous ones. The possibility of polymorphous transitions in
liquid area is under permanent discussion. Another kind of
transitions, i.e. melting, is associated with global order
changes. To describe the phase diagram, one has to describe both
transitions in terms of one and the same statistical model or, in
other words, one has to describe all phases of the system in terms
of one and the same set of order parameters. Such a description
may be realized via local state representation in statistical
mechanics of condensed matter introduced by Patashinski
\cite{ls}. The idea that local state representation may describe the
liquid - liquid phase transition was suggested in \cite{idea}, and
we tried to realize it in the present paper. Here, we give brief
introduction into Patashinski model (section 2), then we perform
calculations for the system with two competing local structures.
The results are applied to the binary system and to the system
with two polymorphous local modifications, such as {\bf Se} and
{\bf S} (section 3). Further applications of the model are
discussed in Conclusion.

\section{Patashinski model.}

Physical background of the model arises from the fact that in
condensed substance the interaction on atomic length-scale is
strong enough to result in a substantial restriction of relative
positions of particles. This is being referred to as the local
order in condensed systems. The size of  strongly correlated
cluster and the nature of local order depends on the temperature,
pressure, and composition of the material. The local ordering
implies that atoms positions in the local cluster  correspond to
some deformed ideal pattern. In close-packed materials, one
supposes several geometrical objects (the fragments of HCP and FCC
lattices and icosahedron) to be such a pattern.  The pattern can
be characterized by  geometric parameters which are invariant
under rotations (coordination lengths and numbers, characteristic
angles in the pattern). These parameters are the same for most
clusters and fix up the type of local order in the material.

To account the limitations imposed on relative positions of
particles by the local order, one describes the system in terms of
local order parameters. Since the concept of local order assumes a
space scale larger than mean interatomic distance, a coarsening of
the description is necessary. The resulting model is then a
simplification that allows one to understand  global ordering
starting from the assumed local order and from the interaction of
local orders in different clusters. The formalism considered was
first offered by A.Patashinski and M.Chertkov in \cite{ls}, and
below in this section we will follow this paper.

As was discussed above,  material is treated as a system of small
clusters of equal size, each cluster being in one of a set of $N$
local structural states. The same set of local states is assumed
for all clusters; a number ~$i=\overline{1,N}$~ is assigned to
each of the states of this set. A state of the whole material is
then described by the field ~$\sigma _i(r)$,~ where $\sigma _i(r)$
is an $N$ - component vector defined by the rule: $\sigma
_i(r)=1$, if the state of the cluster placed at point $r$ has
number $i$, and $\sigma _i(r)=0$ in other case. To proceed with a
statistical theory of the local and global order, one has to find
the non-equilibrium free energy (effective Hamiltonian,
\cite{PATPOKR}) as a function of the configuration $\{\sigma
_i(r)\}$. One expects the effective Hamiltonian $H\{\sigma \}$ to
include independent contributions from different clusters, and
interaction energy that depends on the states of more than one
cluster:
\begin{equation}
\label{h}
H\{\sigma \}=-\sum_{r,r'}\sigma _i(r)M_{ij}(r-r')
\sigma_j(r')-\sum_r\sigma _i(r)\alpha _i(r)
\end{equation}
Here and below, the Einstein's summation rule is supposed. In (\ref{h}), $%
\alpha _i$ is internal energy of a cluster having state number
$i$, and interaction energy is written as two-cluster interaction
of general form. Free energy (\ref{h}) defines the order parameter
field
\begin{eqnarray}
\label{proba}
<\sigma _i(r)>=w_i(r)= \frac{1}{Z}\sum_{\{
\sigma \} }\sigma _i(r)\exp (-\frac{H\{\sigma \}}T), \nonumber\\
Z=\sum_{\{ \sigma \} }\exp (-\frac{H\{\sigma \}}T)
\end{eqnarray}
Here, the partition function $Z(\alpha _i(r), T)$ is a sum over
all local states. As can be seen from (\ref{proba}), the $\omega
_i(r)$ is the probability of state numbered by $i$ to occur at
point $r$, and
\begin{equation}
\label{sum}\sum_{i=1}^N w_i(r)=1.
\end{equation}
Partition function $Z$ may be treated as generating functional for
the order parameter field $w(r)$:
\begin{equation}
\label{Zw}w_i(r)=\frac{\delta \ln Z}{\delta \alpha _i(r)}.
\end{equation}
To calculate $Z$, one uses Hubbard - Stratonovich transformation.
Namely, let us suppose that the quadratic form in Hamiltonian is
positively defined, and its kernel may be written in the form:
\begin{equation}
\label{rr}M_{ij}(r)=E_{ij}J(r).
\end{equation}
(As it will be shown below, this  is not relevant for the mean
field approximation). Introducing real $N$-component
''conjugated'' field $\psi _i(r)$, one  rewrites $Z$ as
\begin{eqnarray}
& Z =\int \prod_{i;r}D\psi_{i}(r)\exp\{-\frac{1}{2T}
\sum_{r,r'}J^{-1}(r-r')E_{ij}\psi_{i}(r) \psi_{j}(r')\}\times
\nonumber \\ & \times
\prod_{i;r}Tr\exp[\frac{1}{T}(E_{ij}\psi_{i}(r)
\sigma_{j}(r)+\alpha_{j}(r)\sigma_{j}(r))]. \label{Z1}
\end{eqnarray}
Here, the integration $D\psi $ goes over all components $\psi _i$
of the field $\psi (r)$ in each point $r$. In (\ref{Z1}), the
summation over local states can be easily done:
\begin{eqnarray}
&Z=\int \prod_{i;r}D\psi_{i}(r)\exp\{-\frac{F}{T}\},\nonumber \\
&F=\frac{1}{2}\sum_{r,r'}J^{-1}(r-r')E_{ij} \psi_{i}(r)
\psi_{j}(r')-\nonumber \\
&-T\sum_{r}\ln[\sum_{i}\exp\{\frac{1}{T}(E_{ij}
\psi_{j}(r)+\alpha_{i}(r))\}]. \label{Z2}
\end{eqnarray}
In this relation, $\psi _i(r)$ is a continuous variable, so one is
able to apply well-known method of analysis - to find out the most
probable configuration which minimizes the thermodynamic
potential, then to investigate  fluctuations around it, etc. Here,
we shall restrict ourselves with the first step, which is the mean
field approximation (MFA).

Function $\psi (r)$, which minimizes the thermodynamic potential
$F$ in (\ref {Z2}) (most probable configuration) obeys following
equation:
\begin{eqnarray}
&\sum_{r'} J^{-1}(r-r') \psi_{i}(r')=Z^{-1}_{0}\exp[\frac{1}{T}
(E_{ij}\psi_{j}(r)+\alpha_{i})],\nonumber \\ &Z_{0}=\sum_{i}
\exp[\frac{1}{T}(E_{ij}\psi_{j}(r)+\alpha_{i})]. \label{eq1}
\end{eqnarray}
Substituting  solution of this equation into (\ref{Z2}), and using
relation $F(\psi )=-T\ln Z$, one gets the generating functional
$Z$. The probability $w_i(r)=-\delta F/\delta \alpha _i(r)$
coincides then with the right side of (\ref{eq1}). Then, for
$w_i(r)$, one has
\begin{eqnarray}
w_{i}(r)=<\sigma_{i}(r)>=\sum_{r'}J^{-1}(r-r') <\psi_{i}(r')>.
\label{w3}
\end{eqnarray}
The $F$ is similar to nonequilibrium thermodynamic potential in
the Landau theory, with order parameter $\psi ^\alpha (r)$. At
equilibrium, $F$ is minimal. Let us define $w_i(r)$ for
nonequilibrium $\psi $
 using (\ref{w3}). Now the nonequilibrium potential $F$ may be
 written in terms of $w_i(r)$:
\begin{eqnarray}
&F=\frac{1}{2}\sum_{r,r'}M_{ij}(r-r') w_{i}(r)w_{j}(r')-\nonumber
\\ &-T\sum_{r}\ln[\sum_{i}\exp(\frac{1}{T}
(\sum_{r'}M_{ij}(r-r')w_{j}(r')+\alpha_{i}))]. \label{Gamma1}
\end{eqnarray}
As can be seen from (\ref{Gamma1}), condition (\ref{rr})  is not
essential in MFA. The nonequilibrium thermodynamic potential
written in terms of order parameter field $w_i(r)$, is the base of
Landau-like theory. At equilibrium, one has
\begin{eqnarray}
\frac{\delta F}{\delta w_{i}(r)}=0, \label{eq2}
\end{eqnarray}
with normalizing condition (\ref{sum}). In explicit form, this
equation may be written as
\begin{eqnarray}
w_{i}(r) = Z^{-1}(r)\exp(\frac{E_{i}(r)}{T}) \nonumber\\ E_{i}(r)
= \sum_{r'} M_{ij}(r-r') w_{j}(r') + \alpha_{i}(r)
\nonumber\\ Z(r)=\sum_{i=1}^{N} \exp(\frac{E_{i}(r)}{T}).
\label{MPA}
\end{eqnarray}
Here, the $E_i(r)$ may be understood as the energy of cluster
having local state number $i$, while the states of surrounding
clusters are
characterized by the mean probability $w_j(r')$. If the field $%
\alpha $ depends no on spatial coordinates, $\alpha (r)=\alpha (0)$, then
$E_i,w_i$ are spatially homogeneous too:
\begin{eqnarray}
E_{i}=M_{ij} w_{j}+\alpha_{i},\ \ \ M_{ij}=\sum_{r}M_{ij}(r).
\label{E1}
\end{eqnarray}
The normalizing condition (\ref{sum}) is fulfilled for solutions
of (\ref{MPA}) automatically. Depending on parameters of the
theory ($M_{ij},\alpha _i,T $), equations (\ref{MPA}) may have a
set of solutions. In the vicinity of the first order phase
transition, one expects at least two minima of $F$, corresponding
to stable and metastable phases. The lowest minima corresponds to
the stable one. As an example of application of formalism
described, let us consider $N$ - local states model with special
interaction:
\begin{eqnarray}
M_{ij} = J \delta_{ij}~, \nonumber\\
 \alpha_{i}(r) = 0
\label{diag}
\end{eqnarray}
which is the Potts model \cite{wu}. In \cite{MP1}, the Potts model
was applied to describe the melting of a single-component
material. The set of local states in this case should be
understood as a set of allowed orientations of local clusters. The
physical picture is that each cluster of condensed material may be
treated as a deformed cluster of ideal crystal. It is obvious for
crystal phase; For liquid, this statement is an assumption
\cite{MP1}, which seems to be correct for a wide range of
materials. For locally crystalline material, one studies the
statistics of orientations of local anisotropy. To obtain the
effective Hamiltonian of orientational degrees of freedom, one
uses following considerations. If two clusters share a boundary,
then the minimum of their interaction energy corresponds to the
coinciding of their orientations. If the minimum is deep enough,
then one is able to divide the space of orientations in equal
cells and to use a coarsened description of orientations in terms
of orientation cells. The number of cells has to be chosen as to
comprise the orientation attraction region, where the interaction
energy is close to its minimum. Denoting the depth of the minimum
as $J$, and the number of cells as $N$, one arrives at the
effective Hamiltonian (\ref{diag}) of the Potts model which
behavior is known. At low temperatures, the material is
orientationally ordered: all clusters with probability ~$\approx
1$~ have one and the same orientation (crystal). At high
temperatures, all orientations are of equal probability (liquid).
The melting of the material is therefore described as an
orientation disordering. More detailed microscopic consideration
of the relation between the Potts model and melting may be found
in \cite{PS}. Note that an appropriate choice of two
phenomenological parameters ~$J,N$~ makes it possible do relate
correctly several quantities, such as melting
 temperature, spinodal temperatures, the hidden heat of transition,
 heat capacity jump.
That is more than one would expect from such a simple model.

The melting theory, based on the Potts model, and our further
considerations essentially deal with the statement that the local
order does not change via melting - the same set of ideal patterns
may be assigned to the liquid and to the crystal. This statement
has no universal character, this is suggestion. In some cases,
inherent amorphous structures may be involved during melting
\cite{inherent}. But for systems considered, there exist
supportive arguments for the crystalline local structure both in
the crystal and in the liquid \cite{b2}.

\section{System with two competing local structures}

In this section, we  discuss an idealized model of a material
having two possible local structures. Each structure has its own
set of allowed orientations. One introduces two indexes to
enumerate the local states:
\begin{equation}
\sigma _k^i;\ \ \ i=1,2;\ \ \ i=1\rightarrow k=\overline{1,n};\ \
\ i=2\rightarrow k=\overline{1,m}.
\end{equation}
The upper index corresponds to the type of local order, while the
lower one enumerates the orientations. The first and the second
structures have $n$ and $m$ allowed orientations respectively. If
only two-cluster interactions are accounted, then Hamiltonian
takes the form
\begin{equation}
\label{H}-H=\alpha \sum_r\sum_{i=1}^n\sigma
_i^1(r)+\sum_{r,r'}\sigma _k^i(r)M_{kl}^{ij}(r-r')\sigma _l^j(r'),
\end{equation}
where $\alpha $ is the difference of internal energies of two
local structures. The kernel of interaction differs from zero only
for nearest neighbors. We suppose it may be written in the form:
\begin{equation}
\label{M}M_{kl}^{11}=\tilde J_1\delta _{kl};\ \ M_{kl}^{22}=\tilde
J_2\delta _{kl};\ \ M_{kl}^{12}=M_{lk}^{21}=\tilde \varepsilon
\end{equation}
Such a form of interaction arises from following. Let the two
structures differ sharply from each other. If two clusters with
different structures share a boundary, then their interaction
energy is ~$\tilde \varepsilon $~ despite of their orientations;
the orientational interaction of neighbors having one and the same
local structure is described by the Potts model. Parameters
~$\tilde J_1,\tilde J_2$~ are the depths of orientational
interaction of structures 1 and 2 respectively, and ~$\tilde
\varepsilon $ - is an interstructural surface energy. Following
notations will be used below:
\begin{equation}
\varepsilon =\tilde \varepsilon \nu ;\ \ \ J_1=\tilde J_1\nu ;\ \
\ J_2=\tilde J_2\nu ,
\end{equation}
were $\nu $ is the number of nearest neighbors.

For the mean - field order parameter, one supposes, in analogue
with the Potts model:
\begin{eqnarray}
w_{1}^{1}=w_{1} ; \ \ \ w_{k \neq 1}^{1} =
\frac{p-w_{1}}{n-1} \nonumber\\ w_{1}^{2} = w_{2} ;
\ \ \ w_{k \neq 1}^{2} = \frac{1-p-w_{2}}{m-1}.
\label{sigma}
\end{eqnarray}
Here, ~$p$~ is the mean probability of the first structure.
Relations (\ref {sigma}) realize the idea of orientational
ordering; the number 1 is assigned to the orientation which is
most probable in the crystalline state. Now one is able to
introduce following {\em \bf classification} of possible phases in
the model:

\begin{enumerate}
\item  $w_1\sim p,\ p\sim 1$~ - crystal with the first type of local
structure;

\item  $w_1=p/n,\ p\sim 1$~ - liquid with the first type of local
structure;

\item  $w_2\sim 1-p,\ p\sim 0$~ - crystal with the second type of
local structure;

\item  $w_2=(1-p)/m,\ p\sim 0$~ - liquid with second type of local
structure;
\end{enumerate}

For the thermodynamic potential (\ref{Gamma1}) per one cluster,
one gets
\begin{eqnarray}\label{osnova}
f=&-&J_2(1-p)^2 \left( \tilde{w}_2-\frac{\tilde{w}_2^2}{2}-
\frac{(1-\tilde{w}_2)^2}{2(m-1)} \right) \nonumber\\
 &-&J_1p^2 \left( \tilde{w}_1-\frac{\tilde{w}_1^2}{2}-
\frac{(1-\tilde{w}_1)^2}{2(n-1)} \right)  \nonumber\\
 &-&\varepsilon p(1-p)+T(1-p)\ln(1-p)+Tp\ln p \nonumber\\
 &+&Tp\ln \tilde{w}_1 +T(1-p)\ln \tilde{w}_2+p\alpha~,
\end{eqnarray}
where the values $\tilde{w}_1=w_1/p~,~~\tilde{w}_2=w_2/(1-p)$ obey
equations
\begin{eqnarray}
\tilde{w}_1&=&\left[1+(n-1)\exp\left(\frac{J_1
p(1-n\tilde{w}_1)}{T(n-1)}\right)\right]^{-1}~, \\
\tilde{w}_2&=&\left[1+(m-1)\exp\left(\frac{J_2
(1-p)(1-m\tilde{w}_2)}{T(m-1)}\right)\right]^{-1}
\end{eqnarray}
With the help of (\ref{osnova}), the phase diagram of the model
can be plotted in MFA for two physically different cases,
considered below.

The first case is binary system in which the second type of local
order occurs due to  presence of  second component, i.e the binary
system  with limited miscibility of the components.  The two types
of local order correspond to the pure substances $A$ and $B$.  The
model can be applied to systems in which the two types of local
order differ sharply (have different groups of local symmetry or
incompatible interatomic distances).  In detail, the calculations
are performed in \cite{PA}, where it was shown that all types of
phase diagrams for binary systems with limited miscibility of the
components (eutectic and monotectic ones) are described correctly.
To demonstrate validity of the model, we calculated the
concentration - temperature phase diagrams for well known binary
systems {\bf Ga-Pb} and {\bf Ag-Cu}. The results are plotted on
fig.1. One can see  some deviations from experimental data
published by Elliott
\cite{elliott}. Determination of model parameters  was made by
 following way.
The parameters are $J_1,n,J_2,m,\alpha,
\varepsilon$. Parameters $J_1,n$ and $J_2,m$ correspond to pure
components and were determined from the heat capacity temperature
dependencies of pure {\bf Ag, Cu, Ga, Pb} \cite{zinovjev}. For
binary system, where the concentration of second component is an
external parameter, the value of $\alpha$ does not play any role:
it should be chosen as to provide the fixed concentration of
mixture. The only parameter to fit known diagram is $\varepsilon$!
Thus, the agreement with the experiment is rather good.

For the {\bf Ag-Cu} system the parameters are : $J_1=1.356, ~n=4$
({\bf Cu} ), $J_2=1.234, ~n=4$ ({\bf Ag}), $\varepsilon=-0.275$.
The {\bf Ag} melting temperature was chosen as the unity. For {\bf
Ga-Pb}, the parameters are $J_1=3.18, ~m=11$ ({\bf Pb}), $J_2=1.6,
n=11$ ({\bf Ga}), $\varepsilon=-1.596$. Temperature unity
corresponds to the {\bf Ga} melting point.

\section{Selenium and sulfur.}

Consider now  the system where atoms rearrangement changing the
type of local order  in small cluster is allowed.  Instead of
binary system, one has the substance with polymorphous
modifications. The thermodynamic potential (\ref{osnova})  may be
applied to plot the phase diagram also. Phase transitions arise,
if one changes the  model parameters - $J_1,J_2,\alpha,
\varepsilon$.
 In physical situation, the change of parameters may
occur under pressure. Here, we describe the pressure - temperature
phase diagrams of Se and S which are known experimentally
\cite{b2}. The coefficients in the model were chosen as to fit the
experiment. The calculations are presented at fig.2, fig.3. For
sulfur, the calculations were done by following way. First, the
parameter $n$ (number of orientations of low - pressure structure)
was determined from the  heat capacity data  for melting at
atmospheric pressure. Since there is no reliable heat capacity
data for high pressure structure, the parameter $m$ was supposed
to be near $4$. We suggest this value because of our experience in
similar calculations: for systems with high density, the number of
orientations tends to two "magic" values, $4$ or $11$. In our
case, the value $4$ provides better correspondence to experiment.
We suppose linear pressure dependence for $J_1,J_2,\alpha$ and no
pressure dependence for other parameters ( $n,m,\varepsilon$ ).
The coefficients in linear dependencies
$J_1=a_1p+b_1,~J_2=a_2p+b_2, \alpha=a_3p+b_3$ were determined from
the "tales" of experimental diagram: $a_1,~b_1$ - from the low
temperature and low pressure line between phases 1 and 2,
$a_3,~b_3$ - from the interphase line near critical point,
$a_2,~b_2$ - from the melting line at high pressure. The central
part of the diagram was then plotted automatically. For sulfur,
the parameters are $n=2.5, ~m=4, ~a_1=0.527, ~b_1=2.433,
~a_2=0.558, ~b_2=3.296,~a_3=-0.094, ~b_3=1.535~,
\varepsilon=-4.506$. One can see good correspondence between
calculations and experimental data. At higher temperatures, there
exist another phase transition into metallic liquid,  but it is
out of our consideration.

For selenium, the calculations were done by similar way. The only
difference is that the position of melting line at high pressure
is not known from experiment, so we tried to reproduce the
interphase line between crystal phases at low temperatures, which
position on the diagram was supposed by V.Brazhkin \cite{b2}. So,
the correspondence to the experiment in the case of {\bf Se}
allows to wish better. The parameters for selenium are  $n=2.5,
~m=4, ~a_1=0.623, ~b_1=2.433, ~a_2=0.443, ~b_2=1.3, ~a_3=-0.226,
~b_3=1.145~,
\varepsilon=-4.133$.

In both cases, temperature scale was chosen as to make the melting
temperature at atmospheric pressure equal to unity.

\section{Conclusion}

The main advantage of the model considered is the fact that all
possible phases are described by one and the same order parameter
- vector of mean probabilities. This gives one possibility to
write out the expression for thermodynamic potential which is
valid for all situations, and thus to investigate the phase
transitions.

In the model, the drastic changes of structure in liquid are
described as the continuation of the line of polymorphous phase
transition into the liquid area, where it terminates in the
critical point. Below critical temperature, the two types of local
order are well distinguished, and phases 2 and 4 (see {\em
\bf classification}) are divided by the line of the first order
phase transitions. At critical temperature, the transition is of
second order. Above it, the local structures are degenerated by
thermal fluctuations. Note, that the well-known phase diagrams at
fig.1 and phase diagrams presented at fig.2, fig.3 are the
diagrams of one and the same model. The only difference is that
for fig.2, fig.3 one uses pressure instead of concentration as a
thermodynamic parameter.

Further application of the model concern an interesting line
corresponding to the equilibrium of liquid and crystal having
different types of local order (phases 1 and 4). We hope the model
to explain ''anomalous'' melting of covalent crystals with diamond
structure ({\bf C,Si,Ge}) and compositions of type
~$A_{III}B_V$~({\bf In-Sb, Ga-Sb})~ with zincblende structure.
Melting of these substances is accompanied with changing of the
type of chemical bonds and sharp changing of local order
(coordination number rises approximately in two times)
\cite{Glazov,Ioffe}. This results in essential increasing of
density and in the semiconductor - metal transition during melting
\cite{Glazov}. At the same time, at high pressures, these
substances transform into metallic crystals with the structure of
white ~{\bf Sn} \cite{Drickamer}. One can say that there exist a
line of  semiconductor - metal transition which  changes its
direction at the crossing point with the line of metal crystal
melting. This point is the point of three - phase equilibrium of
covalent crystal, metallic crystal, and metallic liquid. In
principle, the line of coexistence of crystal and liquid with
different local structures may terminate at lower pressure in the
crossing point with the line of covalent crystal - covalent liquid
transition. Then the line of polymorphous transition should be
plotted into the liquid area and terminated in the critical point.
For substances mentioned, this is doubtful because the
''covalent'' melting should take place at large temperatures
\cite{Mott}. It is more reliable that the line terminates near the
liquid - gas equilibrium, which is out of the validity of the
model due to the destroying of local order at high temperatures.

\section{Acknowledgments}

We thank RFBR for financial support (grant N 02-02-96419), and
collective of High Pressure Physics Institute of Russian Academy
of Sciences (V.Brazhkin, R.Voloshin, S.Popova, V.Ryzhov, A.Lyapin)
for helpful discussions.

\newpage
\begin{figure}[ht]
\begin{center}
\includegraphics[width=0.8\textwidth,clip,viewport=
20 0 260 450]{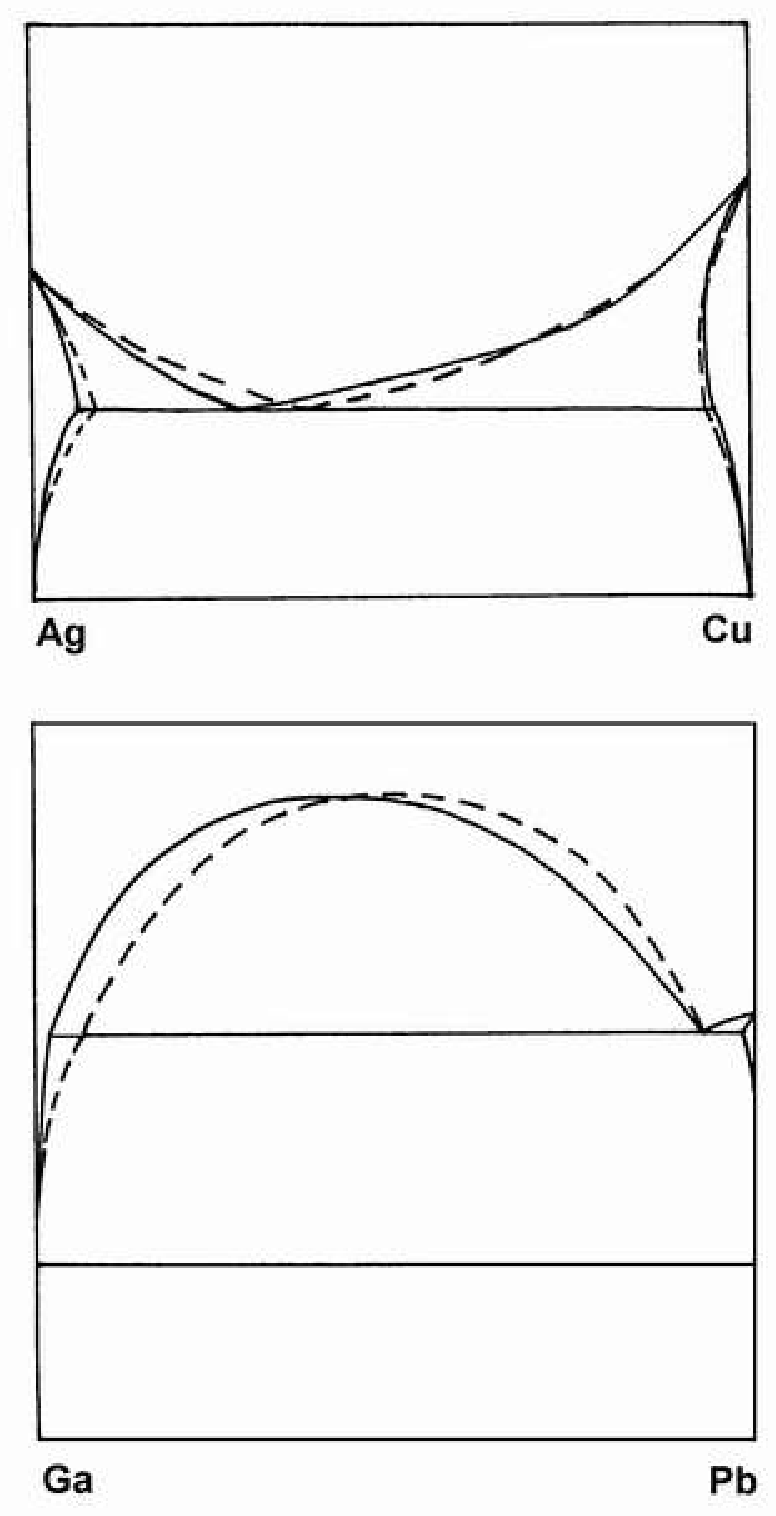}
\label{GaPbAgCu}
\caption{Temperature - concentration phase diagrams of {\bf Ag-Cu}
and {\bf Ga-Pb} binary systems. Solid lines represent experimental
data \cite{elliott}, while the dashed ones correspond to our
calculation.}
\end{center}
\end{figure}

\newpage
\begin{figure}[ht]
\begin{center}
\includegraphics[width=1.0\textwidth,clip,viewport=
20 30 465 345]{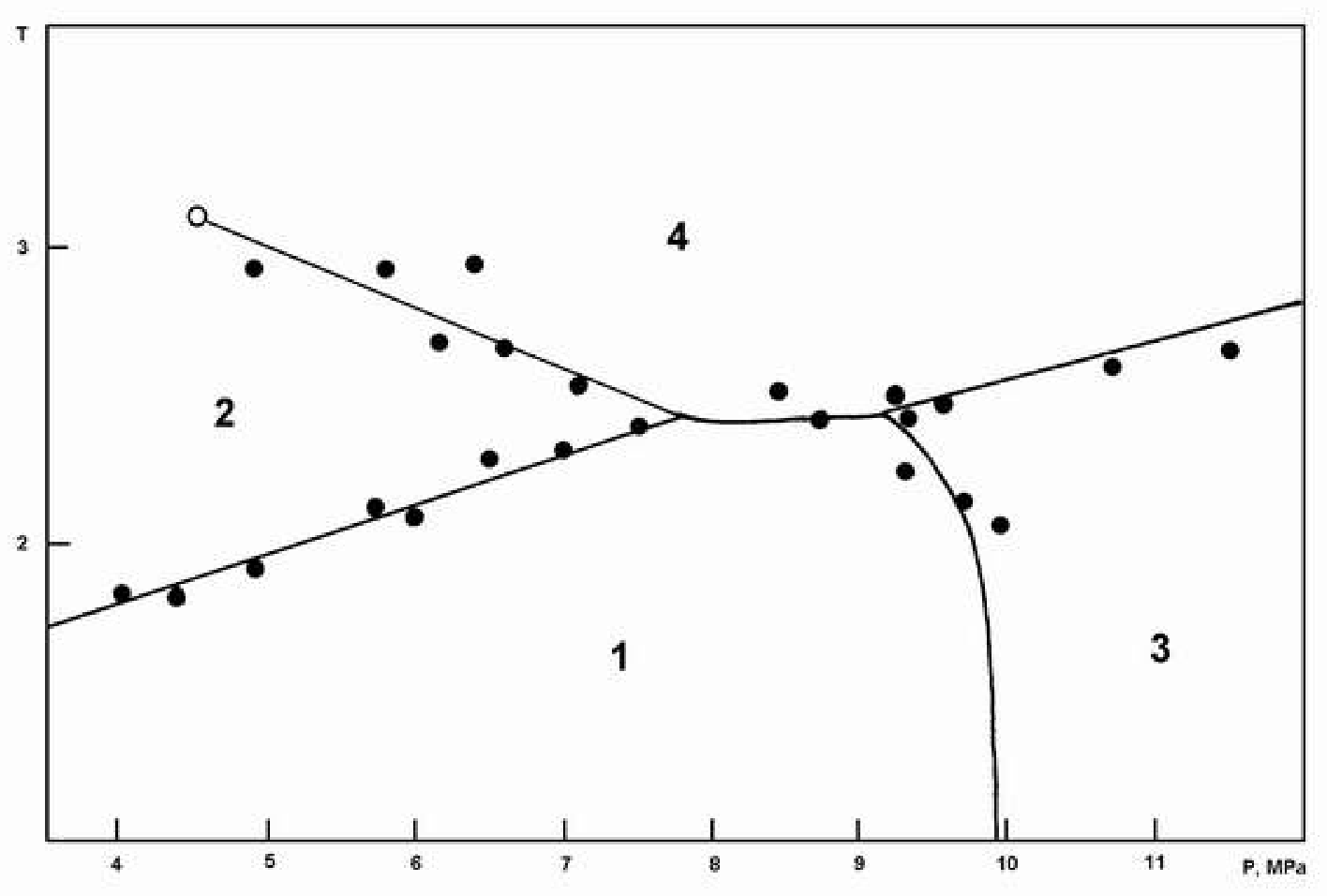}
\label{sera}
\caption{Pressure - temperature phase diagram of sulfur. Solid
circles represent the experimental data, solid lines correspond to
our calculation. Critical point is shown as an empty circle. The
enumeration of phases corresponds to the classification given in
the text.}
\end{center}
\end{figure}

\newpage
\begin{figure}[ht]
\begin{center}
\includegraphics[width=1.0\textwidth,clip,viewport=35 50 430 470]{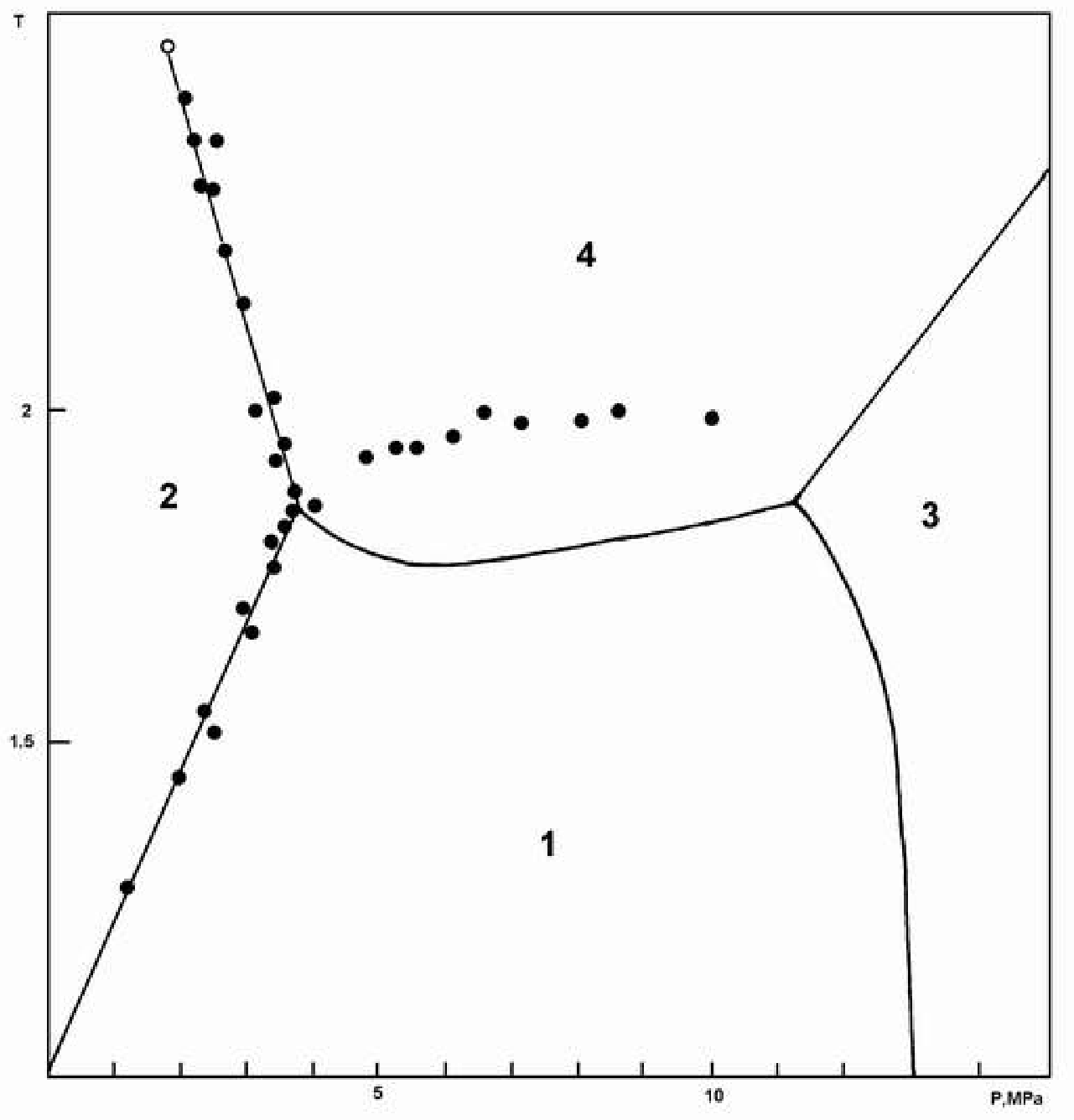}
\label{selen}
\caption{Pressure - temperature phase diagram of selenium. Solid
circles represent the experimental data, solid lines correspond to
our calculation. Critical point is shown as an empty circle. The
enumeration of phases corresponds to the classification given in
the text.}
\end{center}
\end{figure}


\begin{thebibliography}{99}
\bibitem{b1} V.V.Brazhkin, R.N.Voloshin, S.V.Popova, A.G.Umnov, Phys.
Lett.{\bf A 154,} 413 (1991)
\bibitem{b2} V.V.Brazhkin,  S.V.Popova, R.N.Voloshin. Physica {\bf B
265,} 64 (1999)
\bibitem{ls}  A.Z.Patashinski, M.V.Chertkov. Local State Representation in
Statistical Mechanics of Condensed matter. Preprint INP 91-51,
Novosibirsk 1991.
\bibitem{idea} A.Z.Patashinski, A.C.Mitus, B.I.Shumilo. Phys.Lett.
113A, 41 (1985).
\bibitem{PATPOKR}  A.Z.Patashinski, V.L.Pokrovski. {\it Fluctuation theory
of phase transitions}. Pergamon, Oxford, 1979.
\bibitem{wu} F.Y.Wu. Potts model. Rev.Mod.Phys, 54, 239 (1982).
\bibitem{MP1}  A.C.Mitus, A.Z.Patashinski, Sov.Phys. JETP, 53, 798 (1981);
Phys. Lett. 87A, 179 (1982).
\bibitem{PS} A.Z.Patashinski, L.D.Son. Zh.Eksp.Teor.Fiz., 103, 1087 (1993).
\bibitem{inherent} A.Z.Patashinski, A.C.Mitus, M.A.Ratner. Physics
Reports 288, 409 (1997).
\bibitem{PA} A.Z.Patashinski, L.D.Son, G.M.Rusakov, M.A.Ratner.
Physica {\bf A 248} issue 3-4 (1998).
\bibitem{elliott} R.P.Elliott. {\it Constitution of Binary
alloys.} McGraw-Hill, N.Y. 1970.
\bibitem{zinovjev} V.E.Zinovjev {\it Teplofizicheskie svojstva
metallov pri vysokikh temperaturakh}. (Thermophysical properties
of metals at high temperatures) Metallurgy, Moscow, 1989 (in
Russian).
\bibitem{Glazov}  V.M.Glazov, S.N.Chizhevskaya, N.N.Glagoleva, {\it Zhidkie
poluprovodniki}. (Liquid semiconductors) Nauka, Moscow 1967 (in
Russian);
\bibitem{Ioffe}  A.F.Ioffe, A.R.Regel. Progr. Semicond. 4, 237 (1960).
\bibitem{Drickamer}  G.H. Drickamer. in: {\it Phase stability in metals and
alloys}, ed. by P.S.Rudman, Mc-Graw Hill, New York, 1967
\bibitem{Mott}  N.F.Mott, E.A.Davis. {\it Electron Processes in Non -
Crystalline Materials}, Clarendon Press, Oxford, 1979.
\end{thebibliography}
\end{document}